\documentclass[fleqn,usenatbib]{mnras}
\usepackage{graphicx}	
\usepackage{amsmath}	
\usepackage{amssymb}
\usepackage{subcaption}
\usepackage{multirow}
\usepackage{amsmath}
\usepackage{url}
\usepackage{setspace}
\usepackage{longtable}
\usepackage{pdflscape}
\usepackage{float}
\usepackage{booktabs}
\usepackage{tabularx}
\usepackage{amssymb}
\usepackage{wasysym}
\usepackage{rotating}
\usepackage{adjustbox}
\usepackage{xspace}
\usepackage{xcolor}
\usepackage[T1]{fontenc}
\usepackage{adjustbox}
\usepackage{enumitem,calc}

\newcommand{\kms}{kms$^{-1}\,$}

\newcommand{\hi}{H{\sc i}\xspace}

\newcommand\Tstrut{\rule{0pt}{3ex}}         
\newcommand\Bstrut{\rule[-1.5ex]{0pt}{0pt}}   

\title[The first MeerKAT \hi mass function]{MIGHTEE-\hi: The first MeerKAT \hi mass function from an untargeted interferometric survey}

\author[A.~A.~Ponomareva et al.]{Anastasia A.~Ponomareva$^1$\thanks{Email: anastasia.ponomareva@physics.ox.ac.uk}, Matt J. Jarvis$^{1,2}$, Hengxing Pan$^2$, Natasha Maddox$^{3,4}$, Michael G. Jones$^{5}$, 
\newauthor Bradley S. Frank$^{6,7,8}$, Sambatriniaina H. A. Rajohnson$^6$, Wanga Mulaudzi$^{9,6}$, Martin Meyer$^{10}$,
\newauthor Elizabeth A. K. Adams$^{11,12}$, Maarten Baes$^{13}$, Kelley M. Hess$^{14,11}$, Sushma Kurapati$^{6}$, 
\newauthor Isabella Prandoni$^{15}$, Francesco Sinigaglia$^{16,17}$, Kristine Spekkens$^{18}$, Madalina Tudorache$^1$,
\newauthor Ian Heywood$^{1,7,19}$, Jordan D. Collier$^{8,20,21}$ and Srikrishna Sekhar$^{8,22}$\\
$^1$Oxford Astrophysics, Denys Wilkinson Building, University of Oxford, Keble Rd, Oxford, OX1 3RH, UK\\
$^2$Department of Physics and Astronomy, University of the Western Cape, Robert Sobukwe Road, Bellville 7535, South Africa\\
$^3$School of Physics, H.H. Wills Physics Laboratory, Tyndall Avenue, University of Bristol, Bristol, BS8 1TL, UK\\
$^4$Faculty of Physics, Ludwig-Maximilians-Universit{\"a}t, Scheinerstr. 1, 81679 Munich, Germany\\
$^5$Steward Observatory, University of Arizona, 933 North Cherry Avenue, Rm. N204, Tucson, AZ 85721-0065, USA\\
$^6$Department of Astronomy, University of Cape Town, Private Bag X3, Rondebosch 7701, South Africa \\
$^7$South African Radio Astronomy Observatory, 2 Fir Street, Observatory, 7925, South Africa\\
$^8$The Inter-University Institute for Data Intensive Astronomy (IDIA), and University of Cape Town,
Private Bag X3, Rondebosch, 7701,\\ South Africa\\
$^9$Anton Pannekoek Institute for Astronomy, University of Amsterdam, 1098XH, Amsterdam, The Netherlands \\
$^{10}$International Centre for Radio Astronomy Research (ICRAR), The University of Western Australia, 35 Stirling Highway, Crawley,\\ WA 6009, Australia\\
$^{11}$ASTRON, the Netherlands Institute for Radio Astronomy, Oude Hoogeveesedijk 4,7991 PD Dwingeloo, The Netherlands\\
$^{12}$Kapteyn Astronomical Institute, PO Box 800, 9700 AV Groningen, The Netherlands\\
$^{13}$Sterrenkundig Observatorium, Universiteit Gent, Krĳgslaan 281 S9, 9000 Gent, Belgium\\
$^{14}$Instituto de Astrofísica de Andalucía (CSIC), Glorieta de la Astronomía s/n, 18008 Granada, Spain\\
$^{15}$INAF-IRA, Via P. Gobetti 101, 40129, Italy\\
$^{16}$Department of Physics and Astronomy, Universit\'{a} degli Studi di Padova, Vicolo dell'Osservatorio 3, I-35122, Padova, Italy\\
$^{17}$INAF - Osservatorio Astronomico di Padova, Vicolo dell'Osservatorio 5, I-35122, Padova, Italy \\
$^{18}$ Department of Physics and Space Science, Royal Military College of Canada, PO Box 17000, Station Forces, Kingston, K7K 7B4, Canada\\
$^{19}$Department of Physics and Electronics, Rhodes University, PO Box 94, Makhanda, 6140, South Africa\\
$^{20}$School of Science, Western Sydney University, Locked Bag 1797, Penrith, NSW 2751, Australia\\
$^{21}$CSIRO Astronomy and Space Science, PO Box 1130, Bentley, WA, 6102, Australia\\
$^{22}$National Radio Astronomy Observatory, 1003 Lopezville Road, Socorro, NM 87801, USA\\
}

\date{Accepted 2023 April 25. Received 2023 April 21; in original form 2023 March 03}

\pubyear{2023}

\begin{document}
\label{firstpage}
\pagerange{\pageref{firstpage}--\pageref{lastpage}}
\maketitle

\begin{abstract}
We present the first measurement of the \hi mass function (HIMF) using data from MeerKAT, based on 276 direct detections from the MIGHTEE Survey Early Science data covering a period of approximately a billion years ($0 \leq z \leq 0.084 $). This is the first HIMF measured using interferometric data over non-group or cluster field, i.e. a deep blank field. We constrain the parameters of the Schechter function which describes the HIMF with two different methods: $1/\rm V_{\rm max}$ and Modified Maximum Likelihood (MML). We find a low-mass slope $\alpha=-1.29^{+0.37}_{-0.26}$, `knee' mass $\log_{10}(M_{*}/{\rm M_{\odot}}) = 10.07^{+0.24}_{-0.24}$ and normalisation $\log_{10}(\phi_{*}/\rm Mpc^{-3})=-2.34^{+0.32}_{-0.36}$ ($H_0 = 67.4$\,km\,s$^{-1}$\,Mpc$^{-1}$) for $1/\rm V_{\rm max}$ and $\alpha=-1.44^{+0.13}_{-0.10}$, `knee' mass $\log_{10}(M_{*}/{\rm M_{\odot}}) = 10.22^{+0.10}_{-0.13}$ and normalisation $\log_{10}(\phi_{*}/\rm Mpc^{-3})=-2.52^{+0.19}_{-0.14}$ for MML. When using $1/\rm V_{\rm max}$ we find both the low-mass slope and `knee' mass to be consistent within $1\sigma$ with previous studies based on single-dish surveys.
The cosmological mass density of \hi is found to be slightly larger than previously reported: $\Omega_{\rm HI}=5.46^{+0.94}_{-0.99} \times 10^{-4}h^{-1}_{67.4}$ from $1/\rm V_{\rm max}$ and $\Omega_{\rm HI}=6.31^{+0.31}_{-0.31} \times 10^{-4}h^{-1}_{67.4}$ from MML but consistent within the uncertainties. We find no evidence for evolution of the HIMF over the last billion years. 

\end{abstract}


\section{Introduction}
\label{sec:intro}
Hydrogen is the most abundant element in the Universe, constituting around 75 per cent of the total baryonic matter. It is not just the primary building block of all the structure we see in the Universe, but it also plays a crucial role in the formation and evolution of galaxies. Stars in galaxies are born in dense giant molecular clouds, which themselves form due to the cooling of neutral hydrogen. It is unclear, however, where galaxies acquire the fuel to keep forming stars and how star-forming gas gets recycled. Moreover, observational studies show that the rate at which new stars are born in galaxies has been continuously decreasing for the last several billion years \citep[e.g.][]{madau2014,neeleman2016, walter2020}. This decrease must be connected to the amount of the cold gas available to form stars. 

The evolution in the cosmic \hi density ($\rm \Omega_{\rm HI}$) 
is one of the major factors in understanding of the cosmic star formation rate (SFR) density and the mass assembly of galaxies, since \hi serves as a raw material for the build-up of stellar mass \citep{maddox2015, pan2022}. $\rm \Omega_{\rm HI}$ also provides insights into the processes governing the distribution and evolution of cool gas in the Universe. At higher redshifts ($\rm z > 0.2$) $\rm \Omega_{HI}$ has been measured indirectly using either \hi spectral stacking \citep{delhaize2013, bera2018A, rhee2018, Chowdhury2020}, Damped Ly-$\alpha$ (DLA) absorption line systems \citep{peroux2003, noterdaeme2012A&A, grasha2020} or the [C{\sc{ii}}]-to-\hi conversion factor \citep{Heintz2021, Heintz2022}. 
Overall, these studies agree that the \hi mass density of the Universe shows minor evolution with time, in contrast to the molecular hydrogen, which exhibits strong evolution and mirrors that of the global star formation rate density \citep{peroux2020}.

At low redshift $\rm \Omega_{\rm HI}$ is measured directly by summing up the amount of gas in galaxies, usually via determining the \hi mass function (HIMF), which is the neutral hydrogen equivalent of the stellar mass function \citep{baldry2012}. The HIMF defines the number of galaxies per cubic Mpc as a function of \hi mass, and its shape determines how the neutral gas in the Universe is distributed over galaxies of different \hi masses. At $z=0$ the shape of the \hi mass function has been extensively studied \citep{zwaan_himf_2003, martin2010, jones2018, said2019}. These studies have shown that the \hi mass function follows a Schechter function with a power-law low-mass slope ($\alpha$) and an exponential fall-off at the high mass end, beyond a `knee' mass ($M_{\star}$) \citep{zwaan1997}. Even though an agreement between various studies on the faint-end slope of the HIMF has never been reached, they agree that the HIMF depends on the morphological type of galaxies and on the environment where galaxies reside. For example, the overall HIMF tends to have a steeper low-mass slope than when just the Local Group or individual groups of galaxies are considered \citep{ zwaan_env2005, jones_env2016, jones_groups2020, busekool2021}. Therefore, a way to probe an evolution of galaxies over cosmic times is to study variations of the HIMF as a function of morphology, environment and redshift. 

The HIMF is complementary to the stellar mass function in a way that it provides additional insights on galaxies' assembly processes, since the correlation between halo mass and neutral gas mass is very different from that between halo mass and stellar mass \citep{guo2020, yasin2022}. Well-constrained stellar and \hi mass functions put major constraints on the theoretical models of galaxy formation and evolution, as any successful theory of galaxy formation and evolution should be able to reproduce both mass functions simultaneously at any redshift \citep{crain2017, diemer2018, dave2020}. 

To date the most accurate \hi mass function in the Local Universe ($z \leq 0.06$) has been measured by the Arecibo Legacy Fast ALFA survey (ALFALFA; \citealt{martin2010, jones2018}). The resulting HIMF from the ALFALFA 100 per cent survey \citep[][hereafter ALFALFA 100]{jones2018} indicated that most \hi gas in the local Universe resides in the high stellar mass galaxies. Additionally, this study demonstrates the effect of the environment on the HIMF, indicating that the low-mass slope is particularly sensitive. \citet{jones2018} also report a change in the `knee' mass when only part of the sample is used, which is also attributed to the environmental dependence. 
Other observational studies have also shown a flattening of the low-mass slope in high-density environments such as groups and clusters \citep{pisano2011, Westmeier2017, busekool2021}, suggesting that the shape of the HIMF depends on the local and global environment \citep{jones_groups2020}. 

 At redshifts beyond the Local Universe ($z>0.05$), statistical measurements from direct detections in emission become increasingly challenging, due to the intrinsic faintness of the \hi line. However, using associated 21-cm absorption can provide complementary information and push \hi studies to higher redshifts, although these studies are also limited due to the need of strong continuum background sources \citep{gupta2006,maccagni2017, aditya2021}. To date, only two surveys have been able to provide measurements of the HIMF parameters beyond the Local Universe. One is the Arecibo Ultra-Deep Survey (AUDS, \citealt{xi2021}), spanning a redshift range $0 < z < 0.16$. Another is the Blind Ultra-Deep \hi Environmental Survey (BUDHIES, \citealt{gogate2022}), which has constructed the HIMF and measured $\Omega_{\rm HI}$ at $z\sim 0.2$ using direct \hi detections for the first time, but is centred on two galaxy cluster fields. In general, the results of these surveys are in agreement that there is little to no evolution of the \hi content in the Universe up to $z = 0.2$. However, they find a somewhat different low-mass slopes and  `knee' masses. For instance, the results from the AUDS survey are in good agreement with those from ALFALFA 100, especially at the low-mass end, while BUDHIES find a somewhat steeper $\alpha$ and lower $M_{*}$. However, their results are subject to various significant uncertainties, such as completeness corrections and cosmic variance, as well as the environment in which the galaxies reside. \citet{bera2022} have studied the HIMF of star-forming galaxies at $z \sim 0.35$ using stacking technique, and found a significant evolution of the HIMF over the last four Gyr, especially at the high-mass end. In particular, in agreement with BUDHIES, they find a lower `knee' mass and steeper low-mass slope in comparison to the results at $z=0$ from ALFALFA 100. 

The advent of the next-generation deep, blind \hi surveys using new telescopes such as ASKAP \citep{johnston2008}, Apertif \citep{adams2022}, MeerKAT \citep{Jonas_2009} and eventually the SKA will improve our understanding of the redshift evolution of the HIMF and $\rm \Omega_{\hi}$. Surveys such as LADUMA (Looking At the Distant Universe with the MeerKAT Array, \citealt{blyth2016}), DINGO (the Deep Investigation of Neutral Gas Origins, \citealt{meyer2009, rhee2023}) and CHILES (COSMOS \hi Large Extragalactic Survey, \citealt{chiles2019, dodson2022}) are specifically designed to systematically study \hi in galaxies over a large range of redshifts. Another such survey is MIGHTEE (the MeerKAT International GigaHertz Tiered Extragalactic Exploration; \citealt{jarvis2016}), which when complete will detect more than 1000 galaxies in \hi up to $z \sim 0.6$, therefore allowing the systematic study of the evolution of the neutral gas content of galaxies over the past 5 billion years.

In this paper we present the first measurement of the HIMF using data from the MeerKAT telescope. We use the MIGHTEE Early Science data in order to construct the HIMF over the last billion years ($0 \leq z \leq 0.084$) and calculate the cosmic \hi mass density over this redshift range. This is also the first HIMF measured using interferometric data over non-group or cluster field, i.e. a deep
blank field. Our work demonstrates the capabilities of MeerKAT and provides a benchmark for the future \hi evolutionary studies with the SKA pathfinders.

This paper is organised as follows. In Section~\ref{sec:survey} we describe the MIGHTEE Survey and the Early Science data. In Section~\ref{sec:mass} we present the \hi mass measurements and in Section~\ref{sec:himf} we describe how we measure the HIMF. In Section~\ref{sec:results} we present the results and best fit parametrisation of the Schechter function. A summary and conclusions are presented in Section \ref{sec:sum}. 

Throughout this paper, we assume $\Lambda$CDM cosmology parameters of $H_{0}=67.4$\,km\,s$^{-1}{\rm Mpc}^{-1}$, $\Omega_{\rm m}=0.315$ and $\Omega_{\Lambda}=0.685$ \citep{planck2020}.

\begin{figure*} 
\centering
   \includegraphics[scale=0.37]{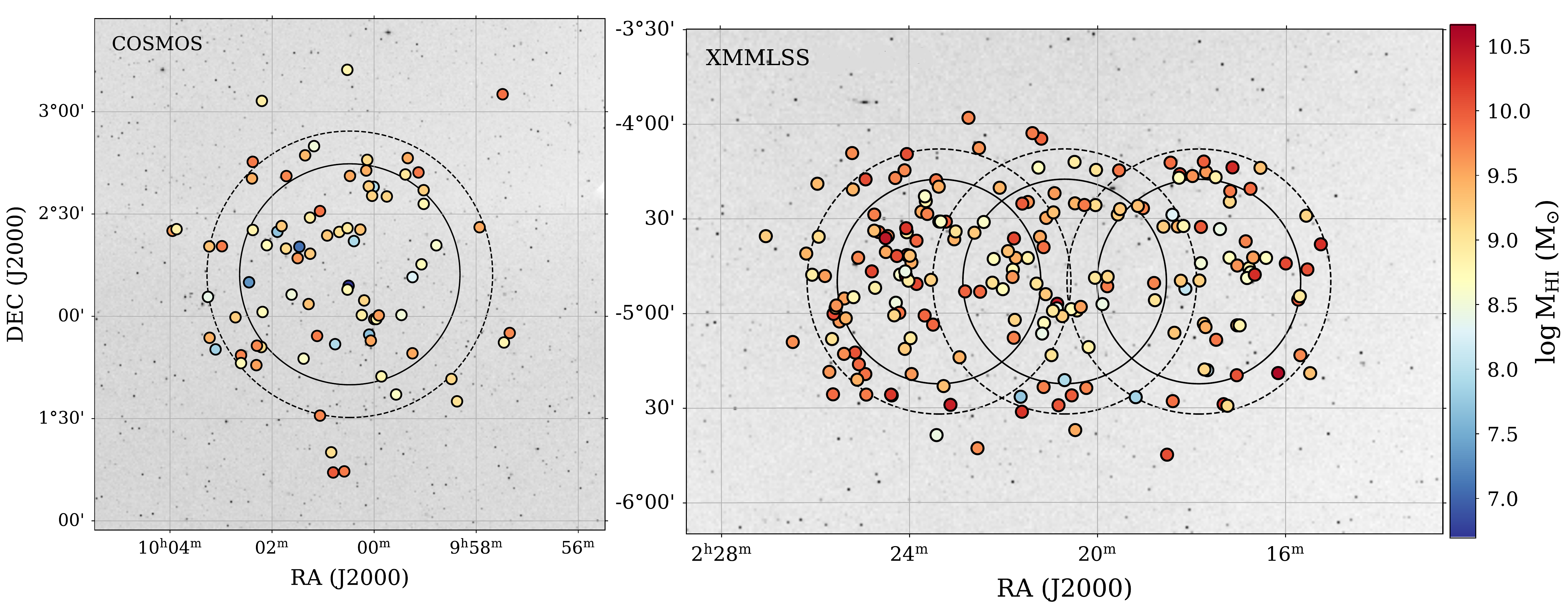}
   \caption{The MIGHTEE-\hi detections in COSMOS (left) and XMMLSS (right) fields colour-coded by their \hi mass. The concentric circles represent the main lobe of the MeerKAT primary beam. The inner solid circle indicates the FWHM of the primary beam $\sim 0.9$ deg$^2$ and the outer 30\% level $\sim $1.5 deg$^2$.
 }
\label{fig_fields}
\end{figure*}

\section{The MIGHTEE Survey}
\label{sec:survey}
MIGHTEE is a MeerKAT survey of four deep, extragalactic fields (COSMOS, XMMLSS, ECDFS, ELAIS-S1) \citep{jarvis2016}. MeerKAT is a radio interferometer that consists of 64 offset Gregorian dishes and equipped with three receivers covering the frequency range from 580 MHz to 3500 MHz \citep{Jonas_2009}. MIGHTEE is simultaneously a spectral line, continuum and polarisation survey. 

For this study we use the Early Science data, which were collected as part of the \hi emission project within the MIGHTEE survey. A detailed description of MIGHTEE-\hi is presented in \citet{maddox2021}. 

The Early Science MIGHTEE observations were conducted between mid-2018 and mid-2019 in L--band ($900 < \nu < 1670$ MHz) with a limited spectral resolution (208 kHz, which corresponds to 44 \kms at $z=0$). A full description of the \hi line data reduction strategy and data quality assessment will be presented in Frank et al. (in prep).

The summary of the Early Science data used in this paper can be found in Table 1 of \citet{sambatra2022}. Briefly, they consist of observations of the COSMOS and XMMLSS fields, and cover the redshift range $0 \leq z \leq 0.084$. The 30 \% area of the main lobe of the primary beam of the COSMOS field is 1.5 deg$^2$ which corresponds to one MeerKAT pointing, although we note that the FWHM of the primary beam is equal to $\sim0.9$ deg$^2$ at 1420.405 MHz. The XMMLSS field was covered by three overlapping pointings resulting in the observed area of 3.3 deg$^2$ (Figure \ref{fig_fields}). The total integration time for the COSMOS field was $\sim17$ hours, and 13 hours were spent on each pointing of the XMMLSS field. The average RMS noise per 208KHz channel across the COSMOS field is $\sigma_{S_{\nu}}=49$\,$\mu$Jy and for the XMMLSS field that average RMS noise is $\sigma_{S_{\nu}}=81$\,$\mu$Jy, corresponding to 3$\sigma$ \hi column density sensitivity of $4.05 \times 10^{19}$ atoms cm$^{-2}$ for COSMOS and $9.83 \times 10^{19}$ atoms cm$^{-2}$ for XMMLSS respectively. 

The source finding was performed visually by inspecting the \hi data cubes with The Cube Analysis and Rendering Tool for Astronomy (CARTA, \citealt{carta}), and was not guided by the available deep optical information. The total Early Science sample consists of 276 objects, each of which has an identified optical counterpart in the very deep multi-wavelength data over these fields. However, we note that a counterpart was not required for a source to be considered genuine.

Although even the full MIGHTEE survey will not be able to compete with the sky coverage of the ALFALFA survey ($\sim$6900 deg$^2$ vs $\sim$32 deg$^2$), the MIGHTEE Early Science flux limit is a factor of ten deeper than the approximate flux limit of the large-area ALFALFA survey, and extends further in redshift (see Figure 8 in \citealt{maddox2021}). For example, a galaxy with an \hi mass equal to $10^9$M$_{\odot}$ and \hi line width equal to 100 \kms would be detectable by ALFALFA out to a distance of $\sim 80$ Mpc \citep{haynes2011,jones2018}, while a galaxy with the same parameters will be detectable by MIGHTEE to $\sim 428$\,Mpc \citep{maddox2021}. This allows us to study both the low mass slope and the `knee' of the HIMF out to larger redshifts. 
 
 The total cosmological volume of the Early Science MIGHTEE survey is $\sim 7000$\,Mpc$^3$. The detections in both fields span approximately the same redshift range from $z_{\rm min}=0.004$ to $z_{\rm max} = 0.082$ in COSMOS and $z_{\rm max} = 0.084$ in XMMLSS, and lie almost exclusively within the area of 1.5 deg$^2$ of each pointing (Figure \ref{fig_fields}).

\section{\hi Mass Measurements}
\label{sec:mass}
The total \hi mass of each galaxy was calculated following the prescription from \citet{meyer2017}:
\begin{equation}
\label{eq:mass}
\left(\frac{M_{\mathrm{HI}}}{{\rm M}_{\odot}}\right) = \frac{2.356 \times 10^{5}}{1+ {z}}\left(\frac{D_{L}}{\mathrm{Mpc}}\right)^{2}\left(\frac{S}{\mathrm{Jy}\,\mathrm{km\,s}^{-1}}\right),
\end{equation}
where $D_L$ is the cosmological luminosity distance to the source, $z$ is redshift and $S$ is the integrated \hi flux density.

The integrated \hi flux density has been calculated using the moment-0 maps constructed individually for each source, taking diffuse low column density emission into account. 
The detailed description of how the moment-0 maps were constructed is presented in \citet{ponomareva2021} and \citet{sambatra2022}. 
The error on the integrated flux $S$ was calculated by projecting the source mask, used to construct the moment-0 map, to four emission free regions around the detection. Then, the uncertainty in the integrated flux of a galaxy was defined as the mean RMS scatter of the four flux measurements in these regions \citep{mpati2016}. As a result, the typical uncertainty on the \hi mass varies from $\sim 5\%$ for the high-mass galaxies to $\sim 20\%$ for the lowest mass objects ($\rm M_{\rm HI} \leq 10^{8}M_{\odot}$). We note that due to MeerKAT's excellent combination of sensitivity and $uv$-plane coverage any missed \hi flux is negligible as compared to the single dish telescopes. For example, we find an excellent agreement between the total fluxes of the overlapping sources from MIGHTEE and ALFALFA (Frank et al., in prep.)

The cosmological distance ($D_L$) to each source has been calculated using the adopted cosmology, following the prescription of \citealt{meyer2017} (Eq.10). 
 According to \citet{tully2014} peculiar velocities are a negligible fraction of observed velocities at $z > 0.03$. In our sample we have 50 objects below this redshift with a mean $\log_{10}(\rm M_{\rm HI} / M_{\odot})=8.3$.
The galaxy with the lowest
systemic velocity of our sample has $V_{\rm sys} = 1238$~\kms, while a typical peculiar motion is $\sim 300$~\kms \citep{darling2018}. Adopting this value as the uncertainty on the systemic velocity results in an uncertainty on $\log_{10}(\rm M_{\rm HI}) \sim 0.06$~dex for the galaxies at $z < 0.03$ and $\sim 0.02$~dex for the whole sample. The overall resulting uncertainties on the H{\sc i} mass due to peculiar velocities are therefore much smaller than the bin size used for our mass function calculation (0.3 dex), and also sub-dominant compared to the Poisson statistics combined with the sample variance (see Section~\ref{sec:var}) for our sample. We therefore do not attempt to correct for such peculiar motion.

\section{Constructing the \hi Mass Function}
\label{sec:himf}
\subsection{Completeness}
\label{sec:compl}
Prior to performing a statistical analysis of any sample, determining the completeness is important. The completeness is usually calculated per \hi mass bin and is an estimate of how many galaxies in that bin have been detected from the  population as a whole given the limitations of the data \citep{gogate2022}.

Many factors can affect whether or not a galaxy is detected in the data. Effects such as 
primary beam attenuation, RFI, non-uniform noise distribution and limited spatial (and/or spectral) resolution all play a role in our ability to detect sources, particularly near the sensitivity limit of the survey. Furthermore, in H{\sc i} surveys in particular, our ability to detect a galaxy depends not only on its total intrinsic flux, but also on its orientation, i.e. galaxies with lower inclinations have higher flux per channel than more inclined galaxies. Low intrinsic flux and high inclinations greatly weaken the possibility of a galaxy to be detected (although the specifics depend on the method adopted for source finding), and therefore \hi galaxy samples tend to be biased towards the most gas rich and low inclination sources in the survey volume. 

The most reliable way to determine the completeness of a survey is to inject artificial but realistic sources into the image cubes and recover them with exactly the same method that was used to find real sources. The recovery rate of the artificial sources as a function of their \hi mass can then be used to correct the underlying HIMF of a survey for completeness. 

This method works well when an automated source finder is available and can be trusted to find sources with high reliability, whilst robustly differentiating real sources from false positives. While there are ongoing efforts to build such source finders for the large \hi interferometric surveys (e.g. \citealt{Westmeier2021}), there is still no definitive solution, especially for the early MeerKAT data with low velocity resolution \citep{healy2021}. For the MIGHTEE Early Science data, we elected to use visual source finding instead of automated methods. A group of people within the MIGHTEE team have examined the data visually, creating source catalogues, which were then merged with duplicates removed. Each source then has been cross-matched with the optical catalogue of each observed field. 
Therefore, we do not perform the completeness correction using injected sources in our data, since it would require the same group of people to repeat the visual source finding on a much larger sample, and it would become prohibitive. However, despite MeerKAT's superb combination of sensitivity and spatial resolution, we will undoutedly be missing sources close to our adopted flux limit, however conversely, the sources we have identified are much more unlikely to be contaminated by artefacts, particular as they have all been cross-identified to counterparts in the exceptionally deep optical \citep{Aihara2019}, and near-infrared \citep{McCracken2012, Jarvis2013} data over these fields \citep[see][for a full description of the combined data set]{adams2022} .

\begin{figure} 
\centering
 \includegraphics[scale=0.7]{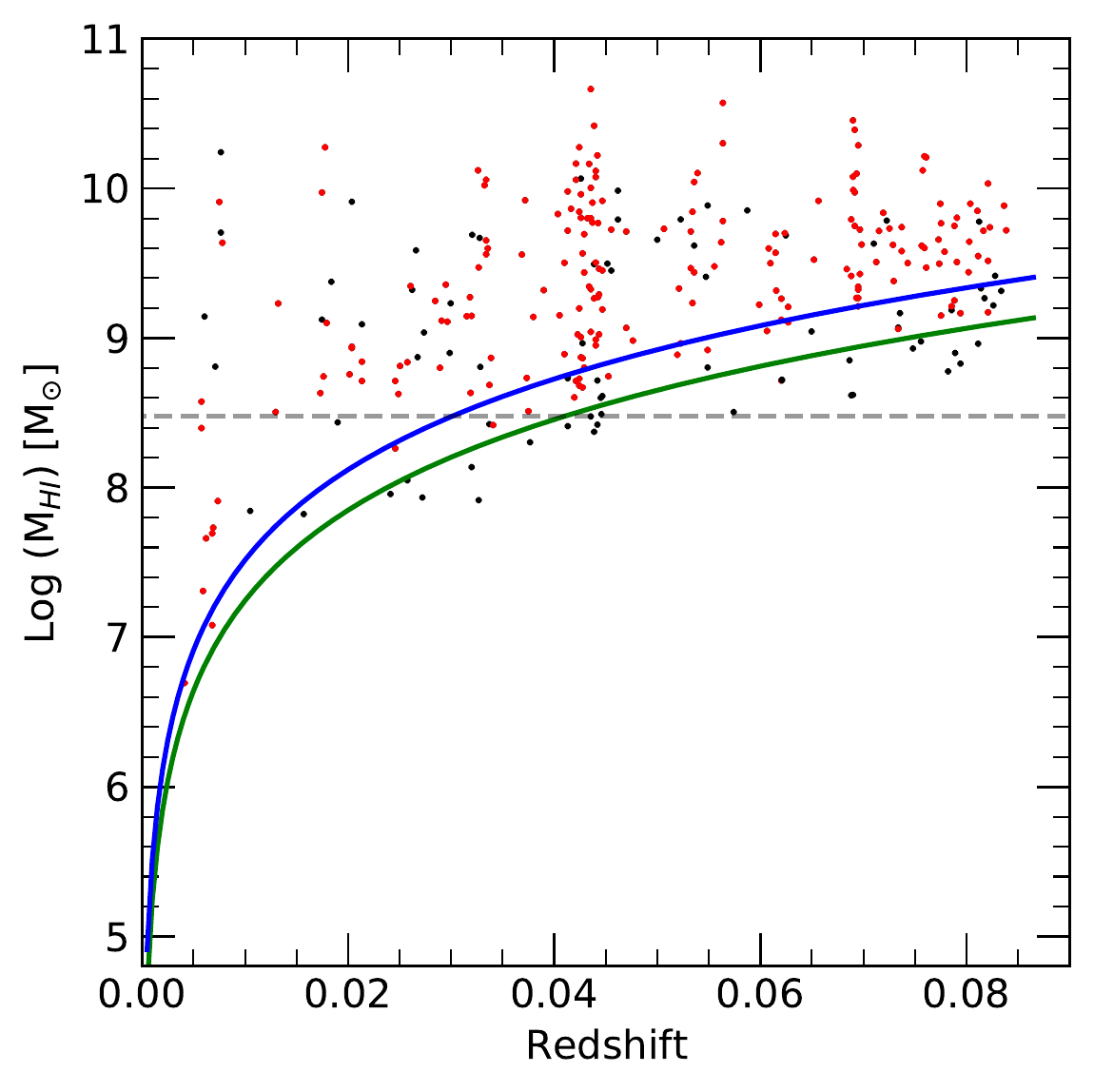}
   \caption{\hi masses as a function of redshift for our sample. Black points indicate the original sample, while red points indicate the sample used to construct the HIMF. The curved green line indicates the $5\sigma$ flux limit used for the simulations of the expected number counts in MIGHTEE-\hi \citep{maddox2021, maddox2016}. The curved blue line represents the median flux limit $5S_{lim}$ (see Section \ref{sec:compl}). The grey dashed horizontal line indicates an additional mass cut-off below which the sample galaxies were discarded (see Section \ref{sec:vmax})}.
\label{fig_MHIz}
\end{figure}

Instead, rather than calculate the incompleteness, we adopt a conservative approach and limit our sample to a flux limit above which we are confident that we are very close to 100 per cent complete. We calculate a limiting line flux density ($S_{\rm lim}$) for each detection as: 
\begin{equation}
\label{eq:lim}
S_{\rm lim}=\sqrt{\frac{W_{50}}{dv}} \times \sigma_{S_{\nu}} \times dv, 
\end{equation}
where $W_{50}$ is the full width half maximum of the \hi line, measured as described in \citet{ponomareva2021}, $dv$ is the velocity resolution at the redshift of the source and $\sigma_{S_{\nu}}$ is the mean measured RMS noise. Since the sensitivity of the telescope decreases with radius from the pointing centre, low-mass galaxies are preferentially detected within the FWHM of the primary beam (as seen in Figure \ref{fig_fields}). Therefore, we measure RMS noise in two areas: within the full-width half power (inner region) and within the 30 per cent of the total primary beam area (outer region). For COSMOS we find that within the  half-power radius $\sigma_{S_{\nu}}=45$$\mu$Jy and at the 30 \% power radius $\sigma_{S_{\nu}}=59$$\mu$Jy. For XMMLSS we find $\sigma_{S_{\nu}}=75$$\mu$Jy in the inner region and  $\sigma_{S_{\nu}}=87$$\mu$Jy in the outer region. To ensure that the sources used to construct the HIMF are detected at least with $5S_{lim}$ independent of their position in the pointing, we remove from the sample all sources with line flux that falls below this limit (based on the relevant field and $\sigma_{S_{\nu}}$ for each source). After the line flux cut and exclusion of the sources detected outside the 1.5 deg$^2$ area (Figure~\ref{fig_fields}) the sample decreases to 203 sources out of 276, which we use to construct the \hi mass function. By adopting this approach we are excluding regions of the survey where we are marginally incomplete to the lower \hi masses, but also not probing as deeply as we potentially could close to the pointing centre. This helps us to mitigate not only redshift-dependent completeness uncertainties, but also completeness uncertainties associated with the position of a galaxy in the field. Figure~\ref{fig_MHIz} shows the distribution of the \hi mass of our sample before and after the flux cut, as well as median flux limit ($5S_{lim}$). To check the robustness of our results with this method to mitigate the incompleteness in our sample, we also adopt a flux limit of 8$S_{lim}$, resulting in a sample of 174 sources. We find that the results are consistent within the uncertainties (Table \ref{tbl_parameters}).

\subsection {$1/V_{\rm max}$ method}
\label{sec:vmax}
The number density of galaxies as a function of their \hi mass can be described as: 
\begin{equation}
\phi(M_{\rm HI})=\frac{dN_{\rm gal}}{dV\,d\log_{10}(M_{\rm HI}) },
\end{equation}
where $dN_{\rm gal}$ is the average number of galaxies in the volume
$dV$, with HI masses that fall within each logarithmic bin in $M_{\rm HI}$.

To date, two different methods have been used in order to convert observed number count of galaxies as a function of their \hi mass into the intrinsic number. One method, the 2-dimensional stepwise maximum likelihood estimator (2DSWML; \citealt{zwaan_env2005,martin2010}), has been widely used for large surveys with high number counts of galaxies per \hi mass bin, as it can incorporate independent methods to account for the effects of the large scale structure (LSS) and sample variance. 

However, if a survey does not have a large number of sources, maximum likelihood methods tend to produce large errors when only a few galaxies per bin are present \citep{zwaan1997, busekool2021}.
Therefore, for our sample we elect to use the so-called $1/V_{\rm max}$ method \citep{schmidt1968}. The principle of this method is to estimate the maximum comoving volume ($V_{\rm max}$) which corresponds to the maximum redshift ($z_{\rm max}$) at which a galaxy of a certain mass could be detected within a given survey. 

To determine $z_{\rm max}$ we use the limiting line flux of each galaxy ($S_{lim}$) calculated in Section \ref{sec:compl}. Then, we iteratively evaluate the line flux ($S$) of each detection over the redshift range of the entire sample (see Figure \ref{fig_MHIz}) until it reaches the value below the detection threshold equal to 5$\times S_{lim}$. The redshift at which this condition is met is assigned as $z_{\rm max}$. We then calculate $V_{\rm max}$ for each detection using $z_{\rm max}$ from the previous step. If $V_{\rm max}$ exceeds the volume corresponding to the upper redshift boundary of the survey volume ($z=0.084$) then $V_{\rm max}$ is set to $V(z=0.084)$. The HIMF is later constructed by summing $1/V_{\rm max}$ in logarithmic bins of \hi mass. 
To further ensure 100 per cent completeness in our sample, we include an additional low-mass cut-off at $M_{H\sc{I}}=3 \times 10^{8}M_{\odot}$. We adopt this cut-off as the most likely source of incompleteness in our sample are low-velocity/low-mass systems where the \hi line-width may only extend over 1-2 channels in our 44 \kms spectral resolution data. Although there are relatively few galaxies below this mass (Figure \ref{fig_MHIz}), we err on the side of caution and only fit the HIMF to galaxies with \hi mass above this limit. Furthermore, the exclusion of these low-mass galaxies does not adversely effect our ability to constrain the low-mass slope of the HIMF, as will be seen in Section~\ref{sec:results}.




\begin{figure*} 
\centering
   \includegraphics[scale=0.8]{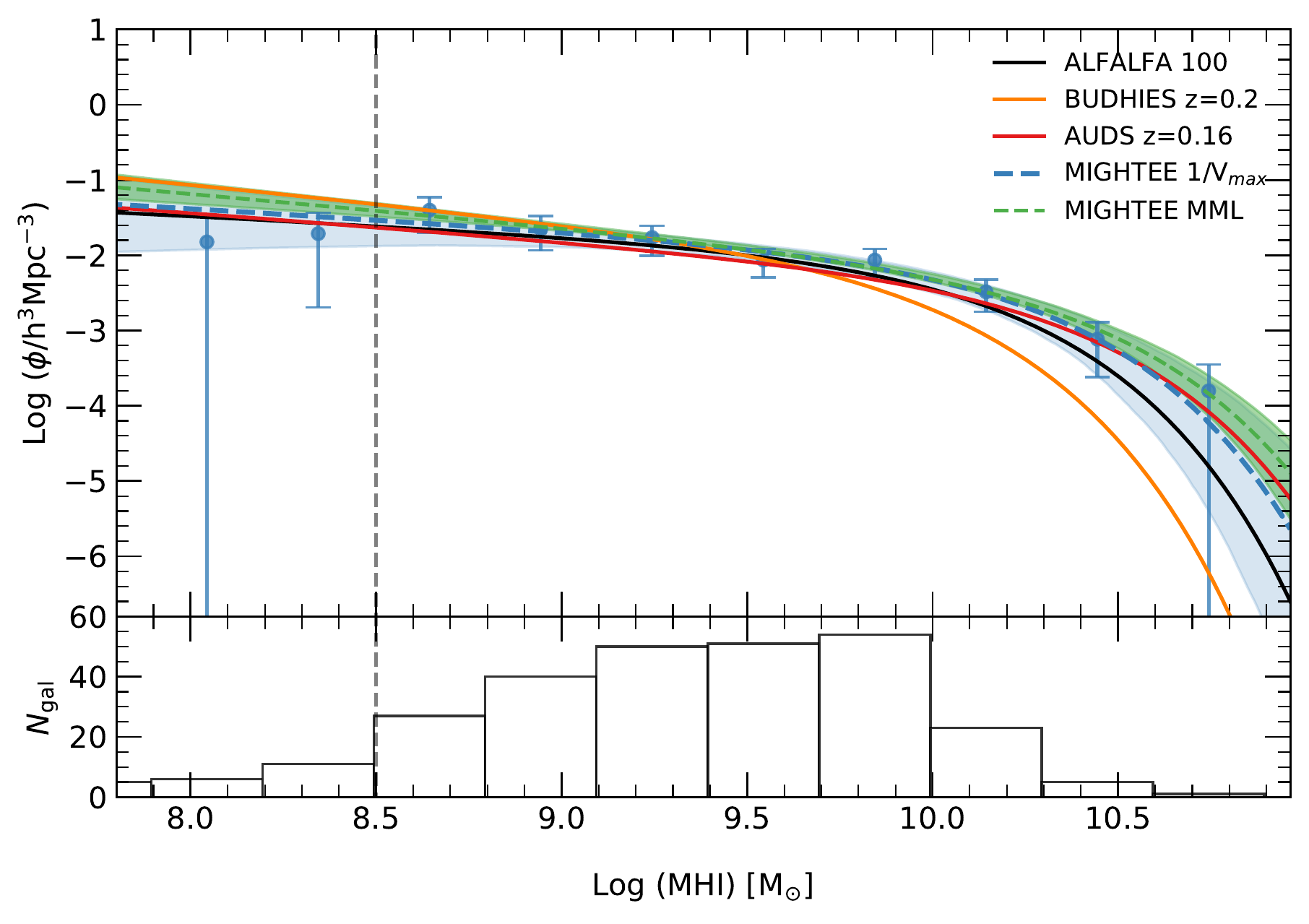}
   \caption{The HI mass function of the MIGHTEE Early Science data is shown with the blue points. The best-fitting relation based on the $1/\rm V_{\rm max}$ method is shown with the blue dashed line.The best-fitting relation based on the MML method is shown with the green dashed line. The ALFALFA 100 HIMF from \citet{jones2018} is shown with the black line. The HIMF measured from BUDHIES \citep{gogate2022} at $z=0.2$ is shown with the orange line, and the HIMF from AUDS \citep{xi2021} at $z=0.16$ is shown with the red line. The histogram in the bottom panel shows the distribution of \hi mass in the MIGHTEE data. The vertical dashed line indicates the mass limit below which the data points were discarded prior to the fit (see Section \ref{sec:compl}). The 1-$\sigma$ uncertainty of the $1/\rm V_{\rm max}$ fit, sampled from the \textsc{Multinest} posteriors (Figure \ref{fig_full_corner}) is shown with the blue shaded area. The 1-$\sigma$ uncertainty of the MML fit sampled from 10$^3$ bootstrap iterations is shown with the green shaded area.}
\label{fig_himf_full}
\end{figure*}
\begin{figure} 
\centering
 \includegraphics[scale=0.46]{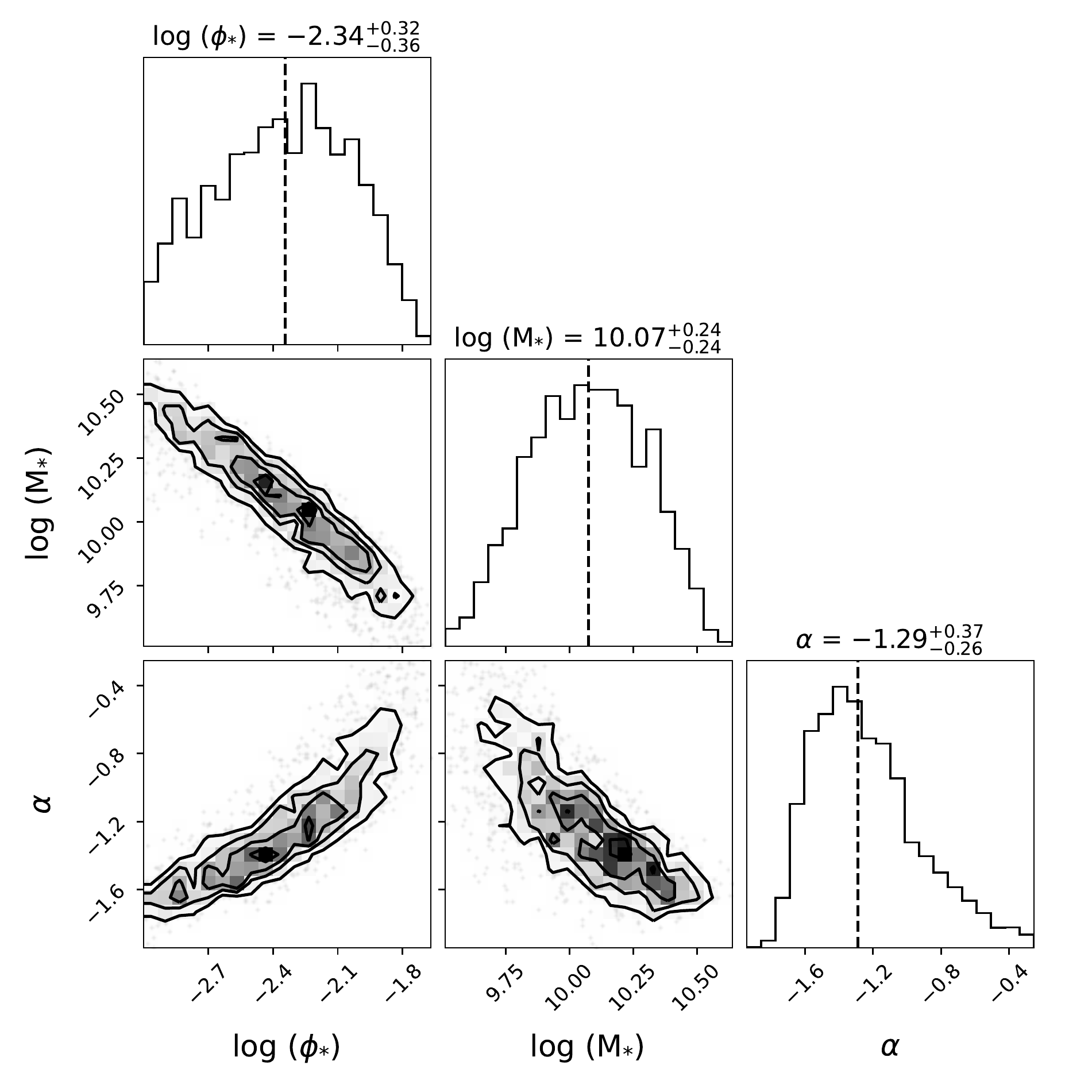}
   \caption{The posterior distributions of the MIGHTEE HIMF parameters ($\phi_{\star}$, $M_{\star}$ and $\alpha$) obtained with \textsc{Multinest}. Black contours on the 2D histograms indicate 1, 2, and 3$\sigma$ confidence levels. Dashed lines on histograms indicate the best-fitting values.}
\label{fig_full_corner}
\end{figure}

\subsection{Sample variance}
\label{sec:var}
Sample variance (sometimes referred to as cosmic variance in this context) is often used to describe the inhomogeneity of the Universe. In other words, the matter in the Universe is not distributed homogeneously and it contains regions of high and low density, which can introduce a systematic bias in observational estimates of the volume density of galaxies. For astronomical surveys that cover large enough volumes and sample all possible environments, sample variance averages out. However, it is a significant source of uncertainty for deep galaxy surveys, which tend to cover relatively small areas. In this case, the mass function tends to be biased by the specific volume and would not be representative of the universal mass function at a particular redshift \citep{somerville2004, moster2011}. 

Calculating the sample variance is not a trivial task. For example, ALFALFA being a large area survey assumed the sample variance uncertainty as the difference between the HIMF of the Spring and Fall skies \citep{jones2018}. 

For MIGHTEE we evaluate the uncertainty due to the sample variance ($\sigma_{v}$) following the prescriptions from the ``Cosmic variance cookbook" by \citet{moster2011}. This prescription uses $\Lambda$CDM predictions to estimate the clustering strength for a given number density at a known average redshift. According to this prescription the sample variance can be estimated by multiplying the dark matter cosmic variance ($\sigma_{dm}$) at a given redshift by the linear galaxy bias ($b$) at that redshift:  
\begin{equation}
\label{cvar}
\sigma_{v}=b(M_{\star}, z)\sigma_{dm}(z,\Delta z=0.1)\sqrt{0.1/\Delta z},
\end{equation}
where $M_{\star}$ is the stellar mass range of the sample, $z$ is the mean redshift of the survey, and $\Delta z$ is the size of the redshift bin. The last term enables a sample variance calculation of different redshift bin sizes. Therefore, the sample variance depends on the stellar mass range within a given sample. Since we have measured stellar masses of our sample galaxies \citep{maddox2021, pan2022}, to evaluate $\sigma_{v}$ we use the stellar mass bins of the MIGHTEE sample. Then we convert stellar mass bins into the equivalent \hi mass bins following the $M_{\rm HI}-M_{\star}$ relation from \citet{maddox2015, maddox2021}, see also \citep{pan2022}. 
\begin{table}
\centering
\caption{The sample variance of the combined MIGHTEE Early Science fields, and for the individual COSMOS and XMMLSS fields, for the full range of \hi mass bins.}
\begin{tabular}{lccc}
\hline
\hline
\multicolumn{4}{c}{Fractional Sample Variance ($\sigma_{v}$)} \Tstrut\\
\hline
log (M$_{\rm HI}$) & MIGHTEE & COSMOS & XMMLSS \Tstrut\\
\hline
6.5 -- 8.5 & 0.20 & 0.36 &0.25\\
8.5 -- 9.5 & 0.22 & 0.38 &0.27\\
9.5 -- 10.0 & 0.24 & 0.41 &0.29\\
10.0 -- 10.5& 0.25 & 0.45 &0.33\\
10.5 -- 11.0& 0.27 & 0.47 &0.34 \Bstrut\\ 
\hline           
\hline
\end{tabular}
\label{tbl_cosvar}
\end{table}

The resulting values for the sample variance of the full MIGHTEE Early Science sample, as well as for the individual COSMOS and XMMLSS fields, are shown in Table \ref{tbl_cosvar}. 
As expected, there is a clear trend and the sample variance decreases with increasing survey area, being the largest for the COSMOS field, and the smallest for the full MIGHTEE sample. As a result, the sample variance introduces the  averaged uncertainties of the volume densities of $\sim 24\%$ for combined fields, $\sim 41\%$ for the COSMOS field and $\sim 30$\% for the XMMLSS field. These uncertainties are in agreement with \citet{driver2010}, who have shown that the survey should be at least ten ``ultradeep'' fields for the effect of sample variance to be below $20\%$ for stellar mass selected samples. We note that \hi-rich galaxies cluster differently than described in Eq.~\ref{cvar} due to the lack of \hi in galaxies in very dense environments \citep{papastergis2013}. Therefore, our sample variance constraints are conservative upper limit estimates. We add these uncertainties in quadrature to the Poisson errors for each \hi mass bin prior to fitting the HIMF.


\subsection{Fitting $1/V_{\rm max}$ data}
\label{sec:fit}
It is widely accepted that a Schechter function can very well describe the shape of the HIMF \citep{zwaan1997}:
\begin{equation}
\label{eq:schechter}
\phi(M_{\rm HI})=\ln (10)\,\phi_{\star}\, (\frac{M_{\rm HI}}{M_{\star}})^{\alpha+1}\, e^{-(\frac{M_{\rm HI}}{M_{\star}})}, 
\end{equation}
where $\phi_{\star}$ is the normalisation constant, $M_{\star}$ is the `knee' mass, and $\alpha$ is the low-mass slope.

In order to determine the best fit we perform a simple $\chi^2$-minimisation, which also enables us to determine the goodness of fit. However, to fully explore the posterior probability distribution of each of the parameters  in the Schechter function, along with their degeneracies, we use \textsc{Multinest} \citep{Feroz2008, Feroz2009}, based on the nested sampling technique \citep{skilling04}. \textsc{Multinest} produces the posterior samples from distributions with an associated error estimate. We use default initial parameters, such as tolerance=0.5 and live points=1000 \citep{Buchner2014}. The prior distributions for $\phi_{\star}$, $M_{\star}$ and $\alpha$ are set in the following ranges: $\rm log_{10}$$(\phi_{\star}$): uniform $\in[-3, 0]$, $\rm log_{10}$$(M_{\star})$: uniform $\in [9.5, 10.5]$ and $\alpha$: uniform $\in [-2.25,  -0.25]$. 
We find that the the maximum likelihood of the best fitting HIMF agrees with the best-fit determined by minimising the $\chi^2$ as expected, but we use the posteriors from \textsc{Multinest} to highlight the degeneracy between parameters. 

\subsection{Modified Maximum Likelihood method}
\label{sec:mml}
As already mentioned above, there is no definitive method to correct small samples of galaxies for the effects of the LSS when measuring the HIMF. While 2DSWML can mitigate these effects for the large surveys, it introduces large errors when there are only a few galaxies per bin of \hi mass present (Section \ref{sec:vmax}). However, to highlight degeneracies associated with fitting an HIMF we use a modified maximum likelihood (MML) method \citep{Obreschkow2018} in addition to $1/\rm V_{\rm max}$. MML method was specifically developed to infer generative distribution functions from uncertain and biased data. This method can accurately recover the mass function of galaxies, while simultaneously dealing with observational uncertainties and to some extent, unknown cosmic large-scale structure. The main difference of this method from $1/\rm V_{\rm max}$ is that it is free of binning and it recovers the shape of the mass function by accounting for the individual $1/V_{\rm max}$ for each galaxy, thus removing the need for binning the data. 
Moreover, it attempts to account for the effects of  LSS (or sample variance) by using the distance distribution of the data to model the mean density of the survey volume at comoving distance $r$ relative to the mean density of the Universe \citep{baldry2012, wright2017}.

For our study we use the R-implementation of the MML (\texttt{dftools}) described in detail in \citet{Obreschkow2018}. For the fit we provide our $1/V_{\rm max}$ values with their associated uncertainties, as well as the distances to our galaxies. To determine the asymmetric uncertainties of the fit we use 1000 bootstrap iterations with a fixed
seed for the random number generator\citep{Obreschkow2018}. The results of this method in comparison to $1/\rm V_{\rm max}$ are presented in Section~\ref{sec:results}, together with comparisons to the literature.

\begin{figure} 
\centering
   \includegraphics[scale=0.55]{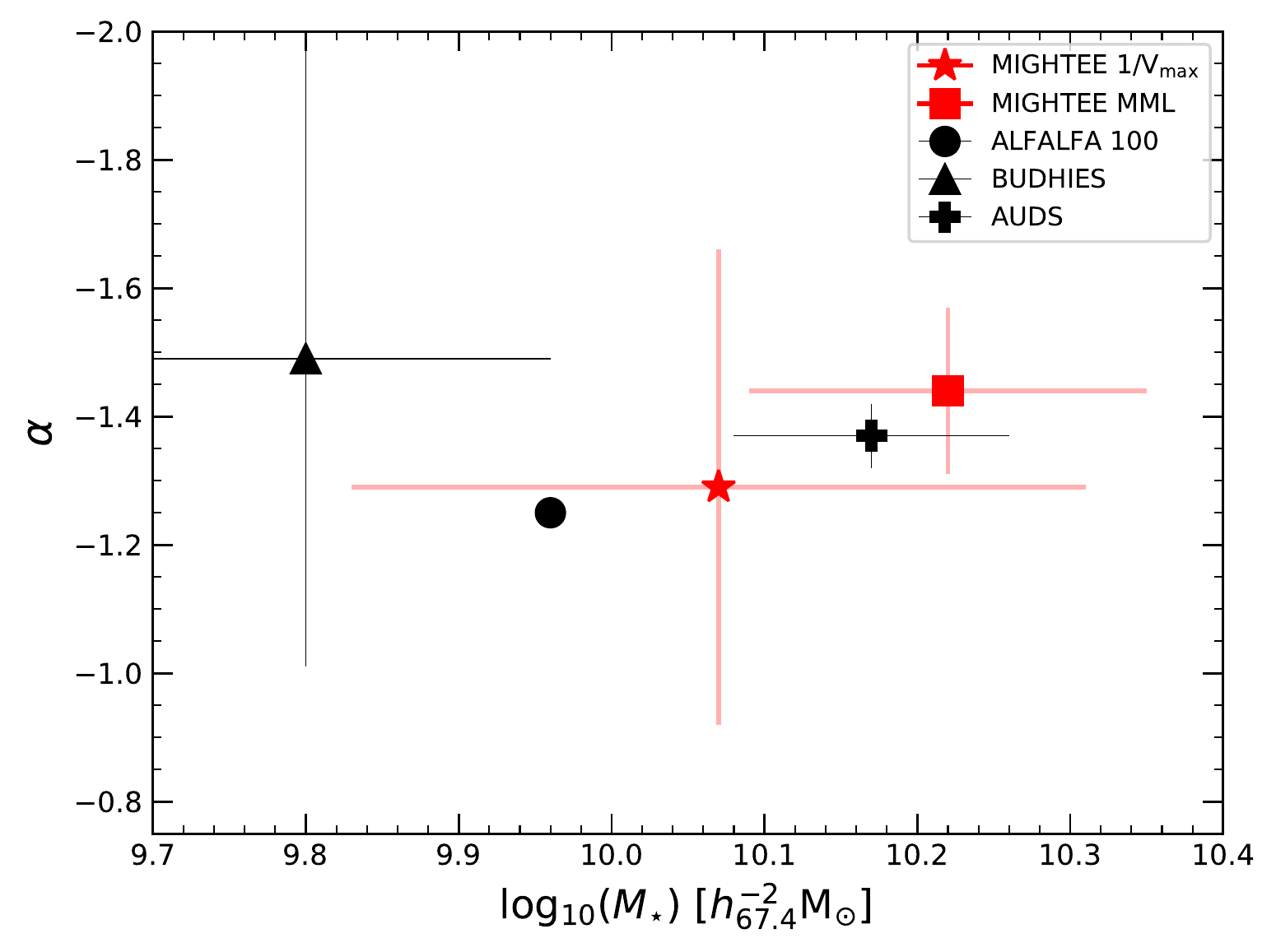}
   \caption{$M_{\star}-\alpha$ measurements with associated uncertainties for different surveys. MIGHTEE measurements are showed with red symbols: star indicates the measurements obtained from $1/ \rm V_{\rm max}$ method and square is from using the MML method (See Section \ref{sec:himf}). The parameters of the three literature surveys are shown with the black symbols and have been scaled to $H_{0}=67.4$\,km\,s$^{-1}{\rm Mpc}^{-1}$ for the ease of comparison.}
\label{fig_param}
\end{figure}

\section{Results}
\label{sec:results}
\subsection{The MIGHTEE HIMF over $0 < z \leq 0.084$ }
Figure~\ref{fig_himf_full} shows the HIMF measured using the MIGHTEE Early Science data together with the best-fitting Schechter function obtained using $1/\rm V_{\rm max}$ method (blue line) and MML method (green line), along with the \hi mass distribution of the sample. The best-fitting parameters ($\phi_{\star}$, $M_{\star}$ and $\alpha$) for both measurements of the Schechter function parametrisation for MIGHTEE and other surveys used for comparison are presented in Table \ref{tbl_parameters}. The posterior distributions for the Schechter function parameters obtained for $1/\rm V_{\rm max}$ method are shown in Figure~\ref{fig_full_corner}.

Overall, the HIMF is very well fit by the Schechter function in both cases, and the results of the two different methods are consistent within the uncertainties. The MML method presents much smaller uncertainties on the parameters since it accounts for the effects of the LSS with an implicit calculation based on the mean galaxy number density. In contrast, for the $1/\rm V_{\rm max}$ method we account for cosmic variance in the error budget of the binned points. For both methods reduced $\chi^2 \approx 1$. The results from both methods are also consistent, within the uncertainties, with the results from ALFALFA 100, AUDS and BUDHIES (Figure \ref{fig_himf_full}).

When comparing the HIMFs from different surveys, samples and redshifts, the most important comparisons arise from the characterisation of the low-mass slope ($\alpha$) and the `knee' mass ($M_{\star}$) since they describe the overall shape of the HIMF. 
For example, using ALFALFA 100 \citet{jones2018} found the low-mass slope to be significantly flatter in the Fall sky than in the Spring sky due to the Virgo cluster, suggesting that $\alpha$ is sensitive to the environment. Using AUDS, \cite{xi2021} found a very similar low-mass slope to ALFALFA 100, even though their sample covers a much larger redshift range ($0 < z \leq 0.16$, as opposed to $0 < z \leq 0.06$ for ALFALFA 100) and a much smaller area.

Using the $1/\rm V_{\rm max}$ method we find the best-fit low-mass slope to be similar to both of these studies, and particularly almost identical to the slope measured by ALFALFA 100 ($\alpha_{A100} =-1.25 (\pm 0.02 \pm 0.1)$ and $\alpha_{\textsc{migthee}} =-1.29^{+0.37}_{-0.26}$). The MML yields a somewhat steeper slope which is consistent with the low-mass slope from BUDHIES, but mainly due to the fact that the low-mass slope from the BUDHIES data is relatively poorly constrained, due to having a mass limit of $M_{\rm HI}=10^9{\rm M}_{\odot}$.
Figure \ref{fig_param} shows the $M_{\star}-\alpha$ comparison between our measurements and those from the literature.

\begin{figure} 
\centering
   \includegraphics[scale=0.73]{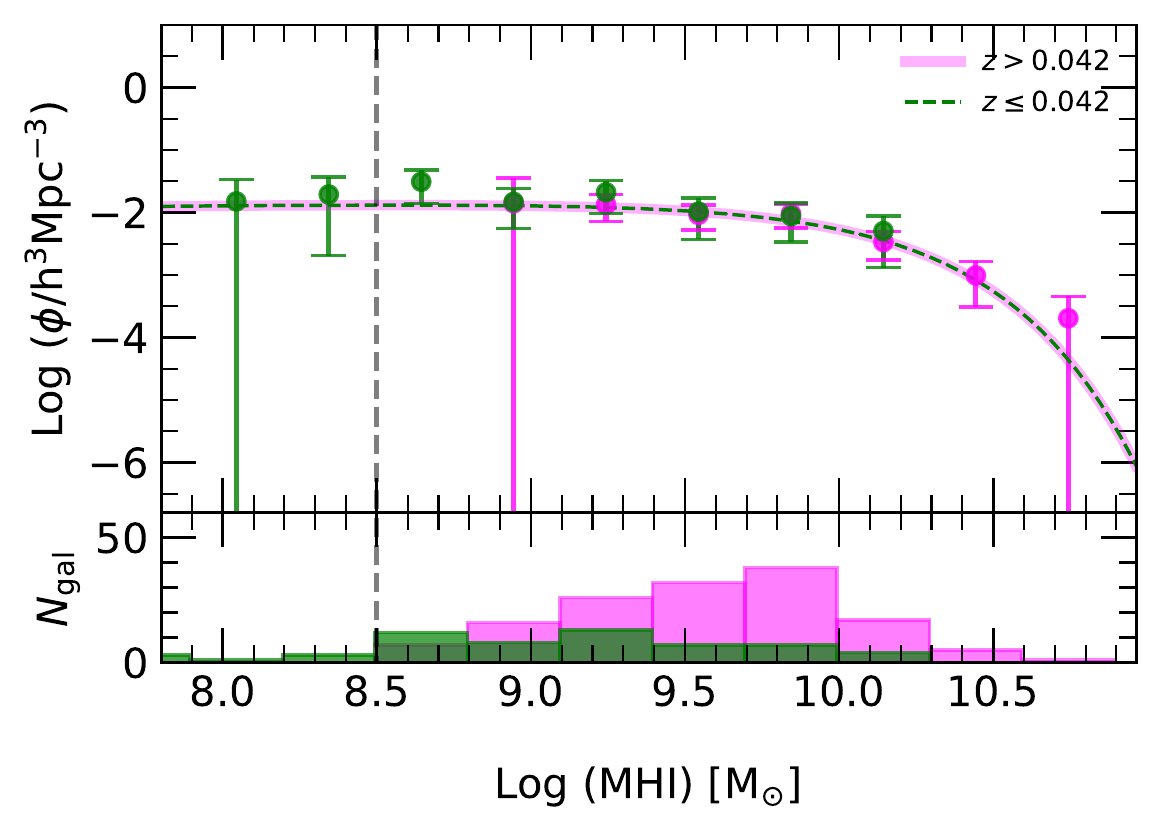}
   \caption{The HI mass function constructed for the evolutionary fit (Eq. \ref{eq:ev}) is shown with the green ($z=0.021$) and magenta lines ($z=0.063$). The green points represent galaxies from the low redshift bin ($z \leq 0.04$), while magenta points are galaxies from the high redshift bin ($z > 0.04$). The lower panel shows the observed counts for `high'-z sample in magenta and low-z sample in green. The vertical dashed line indicates the completeness cut-off.}
\label{fig_himf_z}
\end{figure}

\begin{figure} 
\centering
   \includegraphics[scale=0.36]{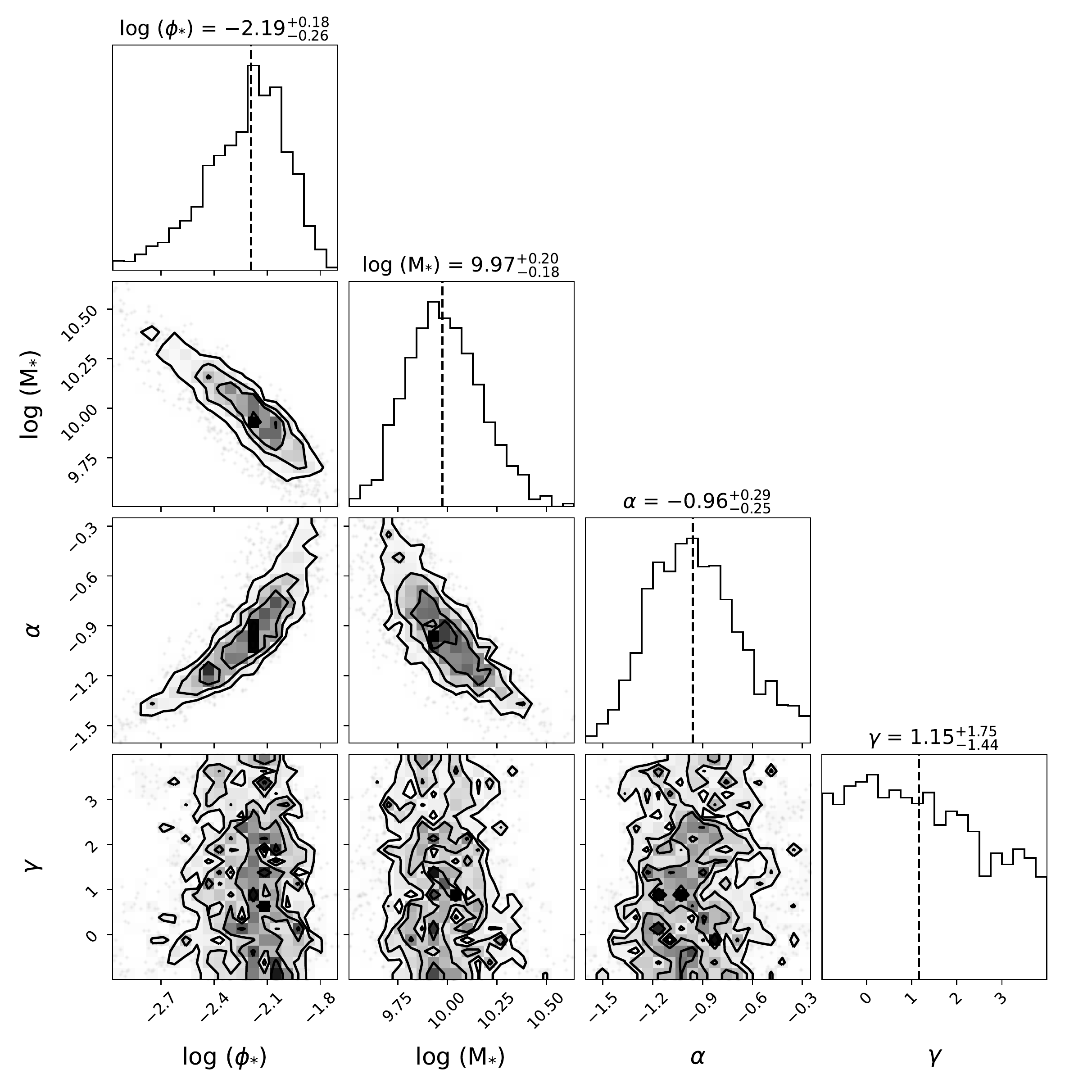}
   \caption{The posterior distributions of the fitted evolutionary modified Schechter function parameters ($\phi_{\star}$, $M_{\star}$, $\alpha$ and $\gamma$) obtained with \textsc{Multinest}. The contours and vertical dashed lines are the same as in Figure \ref{fig_full_corner}.}
\label{fig_z_corner}
\end{figure}

We find the best-fit for the `knee' mass of the HIMF to be consistent, within the uncertainties, with previous studies and is in excellent agreement with AUDS, which probes twice the redshift range of MIGHTEE (Figure \ref{fig_himf_full}). The `knee' mass is responsible for the counts of the high-mass galaxies, which require a large volume to be detected in sizeable numbers. Although, with the MIGHTEE Early Science sample we can constrain the `knee' mass (Figure \ref{fig_full_corner}), the uncertainties are still relatively large (around an order of magnitude larger than ALFALFA 100) due to the limited volume of the Early Science data.

Both our methods ($1/V_{\rm max}$ and MML) yield consistent results (Table \ref{tbl_parameters}). Therefore, we proceed further by comparing only $1/V_{\rm max}$ to the literature studies because it is more sensitive to the uncertainties, and we can investigate the goodness of the fit with the posterior distributions of the Schechter function parameters which highlight possible degeneracies (Figure \ref{fig_full_corner}).

\begin{table*}
\centering
\caption{The best-fitting parameters of a Schechter function parametrisation of the HIMF and resulting $\Omega_{\rm HI}$ for MIGHTEE (this work), ALFALFA 100 \citep{jones2018}, BUDHIES \citep{gogate2022} and AUDS \citep{xi2021}. The parameters of the three literature surveys have been scaled to $H_{0}=67.4$\,km\,s$^{-1}{\rm Mpc}^{-1}$ for the ease of comparison. }
\label{tbl_parameters}
\begin{tabular}{lcccccc}
\hline
\hline \Bstrut\\
Survey (Sample size)&$\rm \log_{10}$($\phi_{\star}$/$h_{67.4}^3$Mpc$^{-3}$) & $\rm \log_{10}$$(M_{\star}$/$h_{67.4}^{-2}\rm M_{\odot}$) &$\alpha$&$\gamma$&$\chi^{2}_{\nu}$&$\Omega_{\rm HI} \times 10^{-4} h_{67.4}^{-1}$ \Bstrut\\
\hline \\
MIGHTEE $1/\rm V_{\rm max}^{5S_{lim}}$ (203)& $-2.34^{+0.32}_{-0.36}$ & $10.07^{+0.24}_{-0.24}$ & $-1.29^{+0.37}_{-0.26}$&-& 0.98 & $5.46^{+0.94}_{-0.99}$ \Bstrut\\
MIGHTEE $1/\rm V_{\rm max}^{8S_{lim}}$ (174)& $-2.36^{+0.35}_{-0.38}$ & $10.10^{+0.22}_{-0.24}$ & $-1.40^{+0.42}_{-0.24}$&-& 1.1 & - \Bstrut\\
MIGHTEE MML (203)& $-2.52^{+0.19}_{-0.14}$ & $10.22^{+0.10}_{-0.13}$ & $-1.44^{+0.13}_{-0.10}$&-& 1.2 & $6.31^{+0.31}_{-0.31}$ \Bstrut\\
MIGHTEE (evolutionary fit) &$-2.19^{+0.18}_{-0.26}$  & $9.97^{+0.2}_{-0.18}$ & $-0.96^{+0.29}_{-0.25}$&$1.15^{+1.75}_{-1.44}$&0.92&- \Bstrut\\
MIGHTEE COSMOS (53)&$-2.93^{+0.29}_{-0.27}$ & $10.16^{+0.46}_{-0.33}$ & $-1.58^{+0.49}_{-0.41}$&-& 2.1 &$2.84^{+1.33}_{-1.05}$ \Bstrut\\
MIGHTEE XMMLSS (150)&$-2.34^{+0.25}_{-0.31}$ & $10.08^{+0.23}_{-0.23}$ & $-1.13^{+0.40}_{-0.27}$&-& 1.2&$4.77^{+0.87}_{-0.92}$ \Bstrut\\
\hline \Tstrut\\
ALFALFA 100 (23621)& $-2.33 (\pm 0.02 \pm 0.07)$ & $9.96 (\pm 0.01 \pm 0.005)$ & $-1.25 (\pm 0.02 \pm 0.1)$&-&-&$4.05 \pm 0.1$ \Bstrut\\
BUDHIES (42)&$-2.30 \pm 0.03$  & $9.80 \pm 0.16$ & $-1.49 \pm 0.48$&-&-&$4.26 \pm 4.6$ \Bstrut\\
AUDS (247)& $-2.60 \pm 0.01$ & $10.17 \pm 0.09$ & $ -1.37 \pm 0.05$&-&-&$3.69 \pm 0.3$\\ 
\hline           
\hline
\end{tabular}
\end{table*}

\subsection{Evolution of the HIMF with redshift}
\label{sec:z}
As a proof of concept we investigate whether or not there is any evidence for evolution in the HIMF as a function of redshift. To do so we divide our main sample into two: a low-redshift sample ($z\leq0.04$) and `high'-redshift sample ($z>0.04$). For both samples we recalculate $V_{\rm max}$ using only the volume corresponding to the new redshift range. We also adjust the Poisson counting errors to the new samples. 

To quantitatively assess the possible evolution of the HIMF, instead of fitting the two samples separately, we rather fit them simultaneously with the modified Schechter function which includes a $(1+z)^{\gamma}$ density evolution term \citep{Pan2020}: 

\begin{equation}
\phi(M_{\rm HI})=\rm ln(10)\,\phi_{\star}\, (\frac{M_{\rm HI}}{M_{\star}})^{\alpha+1}\, e^{-(\frac{M_{\rm HI}}{M_{\star}})}(1+z)^{\gamma}, 
\label{eq:ev}
\end{equation}
where all the parameters are the same as in Eq.~\ref{eq:schechter}, $z$ is the mean redshift of each sample ($z=0.021$ for the low-z sample and $z=0.063$ for the `high'-z sample) and $\gamma$ is an evolutionary parameter describing how the HIMF evolves with redshift. We note that the evolutionary term $(1+z)^{\gamma}$ can be appended to $M_{\star}$ (resulting in a horizonal shift in the HIMF) instead of to the overall density (vertical shift). Given our sample size and where the bulk of our galaxies reside, we have a higher chance of detecting the overall density evolution due to the fact that we have better statistics on the low-mass slope than the high-mass turnover. However, we test this by adopting a characteristic mass-evolution and find similar results, consistent with zero evolution.

Figure~\ref{fig_himf_z} shows the resulting HIMF for both the low and `high'-z samples. From Figure~\ref{fig_himf_z} it is already clear that, we do not detect any significant evolution of the HIMF over the redshift range of our sample. The posterior distributions of fitted parameters are shown in Figure \ref{fig_z_corner}. While all other parameters of the HIMF are well constrained and consistent with the fit of the entire sample (Table \ref{tbl_parameters}), $\gamma$ is not constrained and has very large associated uncertainties (Figure \ref{fig_z_corner}). We calculate $\chi^2_{\nu}=0.92$ using the parameters of the model including $\gamma$ compared to $\chi^2_{\nu}=0.98$ when the evolutionary term is not considered. Therefore, invoking Occam's Razor, within our redshift range we do not find any evidence for evolution of the HIMF. This result is in agreement with various models of galaxy formation and evolution which predict that the major evolution of \hi content of the Universe has occurred between $z=2$ and $z=0$ \citep{yates2021}, and 1 billion years in lookback time is not enough to be able to detect any evolution of the HIMF. 


\begin{figure} 
\centering
   \includegraphics[scale=0.73]{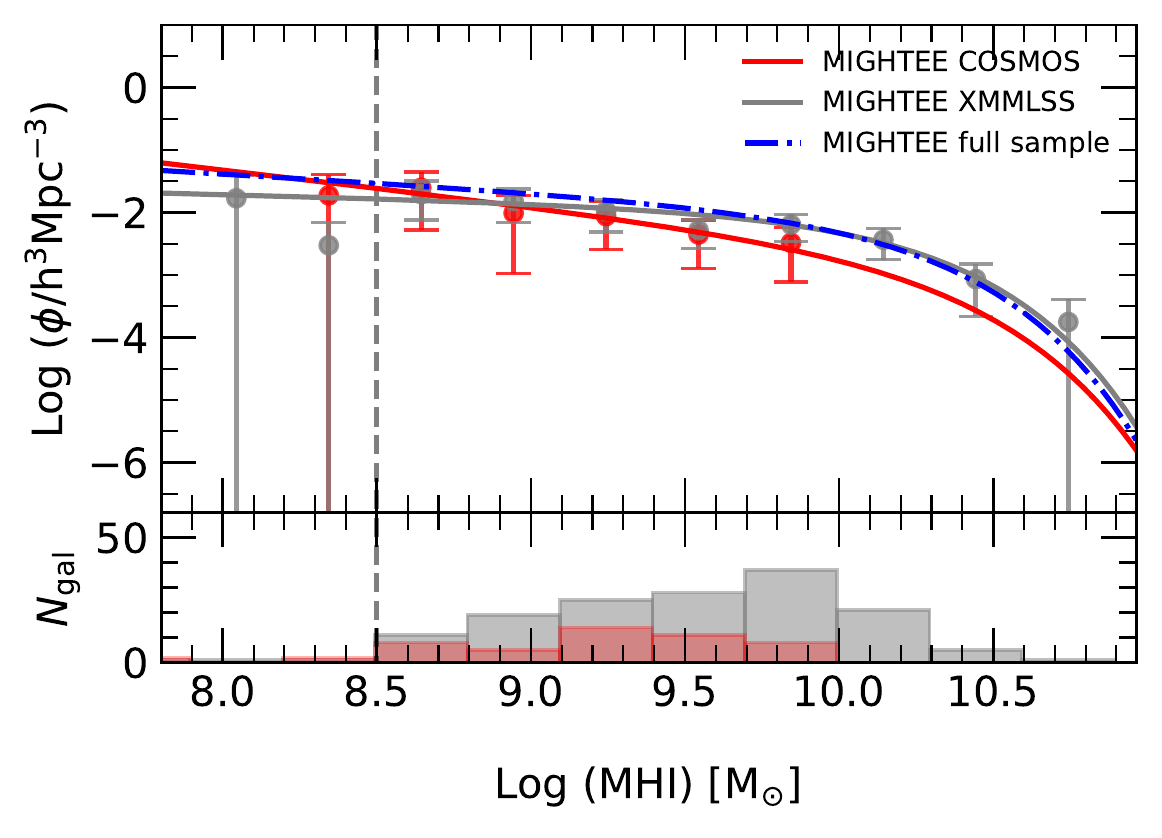}
   \caption{The HI mass function of the COSMOS field is shown with red points and its best-fitting relation is shown with the red line. The XMMLSS HIMF is shown with grey points and grey line. The total MIGHTEE parameterised HIMF is shown with the blue line. The lower panel shows the observed counts for COSMOS in red and for XMMLSS in grey. The
vertical dashed line indicates the completeness cut-off.}
\label{fig_himf_field}
\end{figure}




\subsection{HIMF over different fields}
\label{sec:fields}
Next, we investigate the variations for the HIMF over two distinct fields, which combined make up our main sample, COSMOS (1.5 deg$^2$) and XMMLSS (3.3 deg$^2$). We construct the HIMF for each field separately and adjust the uncertainties due to cosmic variance and Poisson source counts of each sample. The HIMFs for two different fields, together with the HIMF for the main sample are shown in Figure \ref{fig_himf_field}.  

We find that while the HIMF of both fields is well constrained at the low-mass range, 
due to the small volume probed by the COSMOS data, the `knee' mass of the COSMOS HIMF is poorly constrained and suppressed, in comparison to the `knee' mass of the XMMLSS sample. This results in a steeper low-mass slope than when the whole sample is considered (Table \ref{tbl_parameters}). Moreover, the COSMOS HIMF also has much larger uncertainties due to the significantly smaller volume than for the XMMLSS field, which results in lower number counts in the same \hi mass bin coupled with a higher sample variance (Table \ref{tbl_cosvar}).

The `knee' mass of the full sample is completely dominated by the XMMLSS field due to the  much larger volume coverage, and a relatively few high-\hi-mass galaxies in our whole sample have been detected only in the XMMLSS field. Therefore, we do not find any difference between the XMMLSS `knee' mass and the one for the whole sample (Figure \ref{fig_himf_field}).

Due to our limited volume we cannot conclude if the low-mass slope of the HIMF ($\alpha$) is sensitive to the environment, as was found by \citet{jones_groups2020}. 
Given a relatively small sample, we find the measured values of $\alpha$ to have large uncertainties, and therefore the low-mass slopes from the two different fields are consistent with the low-mass slope of the entire sample within errors. Interestingly, $\alpha$ of the main sample sits in between the steeper low-mass slope of the COSMOS field, dominated by low-mass galaxies, and the shallower low-mass slope of the XMMLSS field, dominated by the higher \hi mass galaxies. From Figure \ref{fig_full_corner} and Figure \ref{fig_z_corner} it is clear that the Schechter function parametrisation $\alpha$ and $M_{\star}$ are highly degenerate, and it is only by using the larger volume but slightly shallower XMMLSS data in conjunction with the deeper and narrower COSMOS data, that we can overcome this degeneracy and reduce the uncertainty on each individual parameter.

\begin{figure} 
\centering
   \includegraphics[scale=0.57]{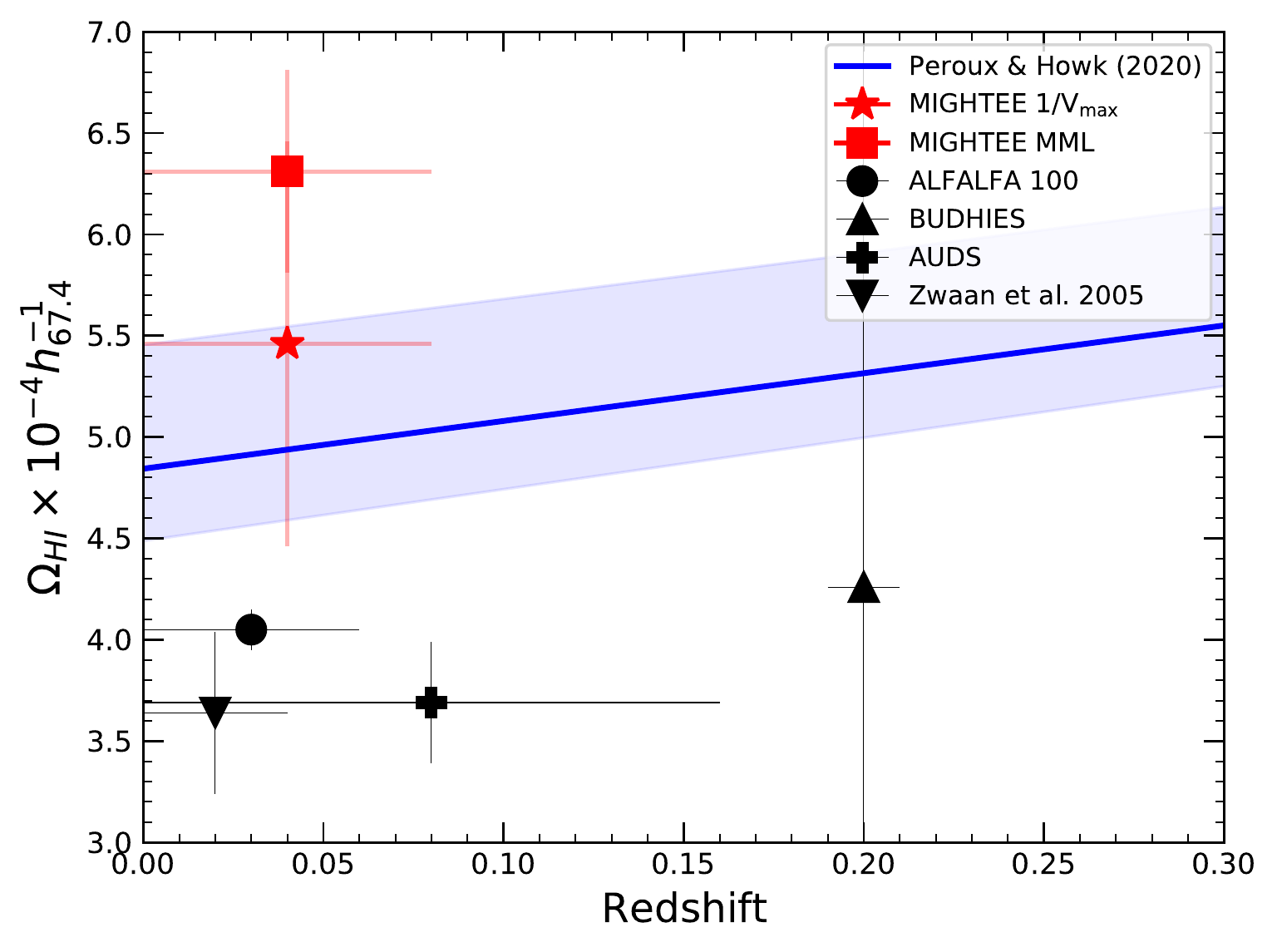}
   \caption{Neutral gas density ($\Omega_{\rm HI}$) as a function of redshift for different \hi surveys (emission only). Horizontal error bars indicate the redshift coverage of each survey. The best-fitting cosmic \hi density from a compilation of \hi emission (direct and stacking) and Ly-$\alpha$ absorption from $z=0$ to $z=5$ is shown with the blue line. The blue shaded region indicates 95\% confidence interval. Adopted from \citet{peroux2020}.}
\label{fig_omega}
\end{figure}


\subsection{Cosmic \hi density ($\Omega_{\rm HI}$)}
The \hi mass function can be used to calculate the cosmic \hi density ($\Omega_{\rm HI}$) by integrating over the best-fitting Schechter function. 

First, the comoving \hi mass density ($\rho_{\rm HI}$) is defined as:
\begin{equation}
\rho_{\rm HI} = \Gamma(\alpha+2)\phi_{\star}M_{*}, 
\end{equation}
where $\Gamma$ is Euler Gamma function and $\phi_{\star}$, $M_{\star}$, $\alpha$ are the HIMF parameters of the best-fitting Schechter function. 
Subsequently $\Omega_{\rm HI}$ is calculated as:
\begin{equation}
\Omega_{\rm HI} = \frac{8 \pi G}{3H_0^2}\rho_{\rm HI}, 
\end{equation}
where $G$ is the gravitational constant and $H_0$ is the Hubble constant.
Using the parameters of the HIMF for our main MIGHTEE sample we obtain $\Omega_{\rm HI}=5.46^{+0.94}_{-0.99} \times 10 ^{-4}$ for the $1/\rm V_{\rm max}$ method and $\Omega_{\rm HI}=6.31^{+0.31}_{-0.31} \times 10 ^{-4}$ using MML method. We remind the readers that in our study we use $H_{0}=67.4$\,km\,s$^{-1}{\rm Mpc}^{-1}$ and therefore the direct comparison with other surveys which use $H_{0}=70$\,km\,s$^{-1}{\rm Mpc}^{-1}$ would not be accurate. If we scale the values of $\Omega_{\rm HI}$ found by the other surveys to $H_{0}=67.4$\,km\,s$^{-1}{\rm Mpc}^{-1}$, we find that our measurements are consistent within 2$\sigma$ with $\Omega_{\rm HI}=4.05 \pm 0.1 \times 10 ^{-4}$ found by ALFALFA 100 \citep{jones2018}. Figure~\ref{fig_omega} shows $\Omega_{\rm HI}$ values measured using direct detections from various \hi surveys scaled to $H_{0}=67.4$\,km\,s$^{-1}{\rm Mpc}^{-1}$ for the ease of comparison. We also show the best-fitting relation for $\Omega_{\rm HI}$ as a function of $z$, obtained by \citet{peroux2020} by fitting the compilation of various $\Omega_{\rm HI}$ measurements from $z=0$ to $z=5$, zoomed in to the relevant redshift range to demonstrate where local measurements lie with respect to the global trend.
Although we find a slightly higher value of $\Omega_{\rm HI}$ using $1/\rm V_{\rm max}$, it is consistent with previous measurements within the large uncertainties, which are dominated by the sample variance, and also with predictions from hydrodynamical cosmological simulations \citep{Villaescusa-Navarro2018, diemer2019} and semi-analytic models \citep{popping2019}. 
For the MML method we find a steeper $\alpha$, higher `knee' mass and similar normalisation, therefore the $\Omega_{\rm HI}$ from MML is larger than when $1/\rm V_{\rm max}$ is used, even though consistent within uncertainties (Figure \ref{fig_omega}).

To investigate how different samples affect the measured $\Omega_{\rm HI}$, we calculate it separately for the two different fields. 
In Section \ref{sec:fields} we found that while COSMOS has a steeper low-mass slope, the XMMLSS field defines the `knee' mass of the entire sample. 
We find $\Omega_{\rm HI}(\rm COSMOS)=2.84 ^{+1.33}_{-1.05} \times 10 ^{-4}$ and $\Omega_{\rm HI}(\rm XMMLS)=4.77 ^{+0.87}_{-0.92} \times 10 ^{-4}$. These are both lower than what we measure for the whole sample due to highly suppressed $M_{*}$ in COSMOS and flatter $\alpha$ in XMMLSS. 

This highlights the importance of fully sampling all parts of the mass function, high-mass galaxies (and therefore large enough volume) are needed to constrain the `knee' of the mass function, whereas the lower mass galaxies are needed to decouple the low-mass slope from variations in $\phi_*$.

\section{Summary and Conclusions}
\label{sec:sum}
In this paper we present the  first \hi mass function determined from data using the MeerKAT telescope and the first using interferometry data over a non-targeted overdensity. We use the Early Science data from the MIGHTEE survey and construct the HIMF over the last billion years ($0 \leq z \leq 0.084$). We investigate the properties of the HIMF in different fields, as well the possible evolution with redshift. Our main results can be summarised as follows:

\begin{itemize}
\item Visual source finding is still widely used to assess the reliability of the detected sources. However, moving forward it is not sustainable for when the modern surveys will reach their full capacity. 
\item We use two different methods to measure the parameters of the HIMF ($1/\rm V_{\rm max}$ and MML) and we find that the first MeerKAT HIMF is in excellent agreement with previous single-dish and interferometric studies, even though all these studies probe different area, volume, environment and redshift range. Using $1/\rm V_{\rm max}$ we find an identical low-mass slope $\alpha=-1.29^{+0.37}_{-0.26}$ in comparison to findings by ALFALFA 100 ($\alpha=-1.25 \pm 0.02\pm 0.01$) and `knee' mass ($\rm \log_{10}(M_{*})=10.07 \pm 0.24$) that is consistent within the uncertainties when comparing to the ALFALFA 100 survey ($\rm \log_{10}(M_{*})=9.96\pm 0.01\pm 0.005$), as well as when comparing to the higher redshift ($z=0.16$) AUDS ($\rm log(M_{*})=10.17\pm 0.09$). We note that we scale the parameters of the literature values to $H_{0}=67.4$\,km\,s$^{-1}{\rm Mpc}^{-1}$ for the ease of comparison.

\item As a proof of concept we investigate whether or not there is any evidence for evolution in the HIMF as a function of redshift. As expected from cosmological models of galaxy formation and evolution, we find no evidence for evolution of the HIMF over the last billion years. This result is also consistent with the studies of the HIMF at higher redshift range -- neither AUDS ($z=0.16$) nor BUDHIES ($z=0.2$) have found any evidence for the evolution of the HIMF. However, the evolution in the shape of the HIMF (suppressed `knee' mass and steeper low-mass slope) is expected at  $z \sim 0.35$ according to findings by \citet{bera2022}. This will be tested with the full capacity of the MIGHTEE survey.

\item We investigate the properties of the HIMF in two distinct fields (COSMOS and XMMLSS). We find $\alpha$ to be steeper in COSMOS due to the suppressed `knee' mass. We find that $M_{*}$ is highly sensitive to the volume observed by a survey, as it requires a sampling of galaxies beyond the `knee', and it is poorly constrained for COSMOS which covers a smaller volume than XMMLSS. 

\item We find the cosmic \hi density $\Omega_{\rm HI}=5.46^{+0.94}_{-0.99} \times 10 ^{-4}$ to be slightly higher than reported by previous studies, though consistent within uncertainties. We find $\Omega_{\rm HI}(\rm COSMOS)=2.84 ^{+1.33}_{-1.05} \times 10 ^{-4}$ and $\Omega_{\rm HI}(\rm XMMLSS)=4.77 ^{+0.87}_{-0.92} \times 10 ^{-4}$, which highlights the importance of fully sampling all parts of the mass function,  large enough volume is needed to constrain the `knee' of the mass function, whereas the depth is needed to constrain the low-mass slope. 
 
\end{itemize}

New observational facilities such as MeerKAT, have the potential to transform our knowledge of \hi in the Universe way before the SKA era. Even with the Early Science MIGHTEE data we were able to measure the HIMF and estimate $\Omega_{\rm HI}$ over the period of the last billion years. At the same time, the MIGHTEE Large Survey Program is well underway, and will give us an opportunity to extend the current
study up to $z=0.5$. With the wealth of excellent ancillary data, including large 4MOST spectroscopic survey \citep{WAVES}, the MIGHTEE data will be crucial for our understanding of the evolution of the HIMF and $\Omega_{HI}$, especially using the combination of direct detections and new Bayesian stacking techniques \citep{Pan2020}.

\section*{acknowledgements}
AAP, MJJ and IH acknowledge support of the STFC consolidated grant [ST/S000488/1] and  [ST/W000903/1] and MJJ and IH support from a UKRI Frontiers Research Grant [EP/X026639/1].
SHAR is supported by the South African Research Chairs Initiative of the Department 
of Science and Technology and National Research Foundation.
MJJ, HP and IH acknowledge support from the South African Radio Astronomy Observatory (SARAO) which is a facility of the National Research Foundation (NRF), an agency of the Department of Science and Innovation.
AAP and MJJ acknowledge support from the Oxford Hintze Centre for Astrophysical 
Surveys which is funded through generous support from the Hintze Family Charitable Foundation. 
NM acknowledges support of the LMU Faculty of Physics.
IP acknowledges support from INAF under the Large Grant 2022 funding scheme (project "MeerKAT and LOFAR Team up: a Unique Radio Window on Galaxy/AGN co-Evolution”.
SK is supported by the South AfricanResearch Chairs Initiative of the Department of Science andTechnology and National Research Foundation.
KMH acknowledges financial support from the grant CEX2021-001131-S funded by MCIN/AEI/ 10.13039/501100011033, from the coordination of the participation in SKA-SPAIN, funded by the Ministry of Science and Innovation (MCIN) and from grant  PID2021-123930OB-C21 funded by MCIN/AEI/ 10.13039/501100011033, by “ERDF A way of making Europe” and by the "European Union".
The MeerKAT telescope is operated by the
South African Radio Astronomy Observatory, which is a facility of
the National Research Foundation, an agency of the Department of
Science and Innovation. We acknowledge use of the Inter-University
Institute for Data Intensive Astronomy (IDIA) data intensive research
cloud for data processing. IDIA is a South African university partnership
 involving the University of Cape Town, the University of Pretoria
and the University of the Western Cape. The authors acknowledge the
Centre for High Performance Computing (CHPC), South Africa, for
providing computational resources to this research project. This work
is based on data products from observations made with ESO Telescopes 
at the La Silla Paranal Observatory under ESO programme
ID 179.A-2005 (Ultra-VISTA) and ID 179.A- 2006 (VIDEO) and on
data products produced by CALET and the Cambridge Astronomy
Survey Unit on behalf of the Ultra-VISTA and VIDEO consortia.
Based on observations collected at the European Southern Observatory
under ESO programmes 179.A-2005 (UltraVISTA), and 179.A-2006 (VIDEO).
Based on observations obtained with MegaPrime/MegaCam, a joint
project of CFHT and CEA/IRFU, at the Canada-France-Hawaii Telescope 
(CFHT) which is operated by the National Research Council
(NRC) of Canada, the Institut National des Science de l’Univers of
the Centre National de la Recherche Scientifique (CNRS) of France,
and the University of Hawaii. This work is based in part on data products 
produced at Terapix available at the Canadian Astronomy Data
Centre as part of the Canada-France-Hawaii Telescope Legacy Survey,
 a collaborative project of NRC and CNRS.The Hyper SuprimeCam (HSC) 
 collaboration includes the astronomical communities of
Japan and Taiwan, and Princeton University. The HSC instrumentation and 
software were developed by the National Astronomical
Observatory of Japan (NAOJ), the Kavli Institute for the Physics
and Mathematics of the Universe (Kavli IPMU), the University of
Tokyo, the High Energy Accelerator Research Organization (KEK),
the Academia Sinica Institute for Astronomy and Astrophysics in Taiwan 
(ASIAA), and Princeton University. Funding was contributed by
the FIRST program from Japanese Cabinet Office, the Ministry of Education, 
Culture, Sports, Science and Technology (MEXT), the Japan
Society for the Promotion of Science (JSPS), Japan Science and Technology 
Agency (JST), the Toray Science Foundation, NAOJ, Kavli
IPMU, KEK, ASIAA, and Princeton University.


This research has made use of NASA’s Astrophysics Data System Bibliographic Services. This research
made use of Astropy\footnote{\href{http://www.astropy.org}{http://www.astropy.org}}
a community-developed core Python package for Astronomy.

\section*{Data availability}
The MIGHTEE Early Science spectral cubes are available by request to the corresponding author. The data catalogue which was used in this paper can be found as supplementary material  (\href{https://www.dropbox.com/s/vyc279ilhd46l7e/HIMF_catalogue.txt?dl=0}{https://shorturl.at/bglpJ}). It includes ID, coordinates, observed frequency, total \hi flux density and its associated error (see Section \ref{sec:mass} for details) for every detected source.

\bibliographystyle{mnras}
\bibstyle{mnras}
\bibliography{MIGHTEE_HIMF}
\vspace*{1cm}

\end{document}